\documentclass[twocolumn,amsmath,amssymb,superscriptaddress,prb]{revtex4-1}

\usepackage{graphicx}
\usepackage{dcolumn}
\usepackage{bm}
\usepackage[abs]{overpic}
\usepackage{here}
\usepackage{color}
\usepackage{url}
\usepackage{accents}
\def\up{\uparrow}
\def\down{\downarrow }
\def\Vec#1{\bm{#1}}

\begin{document}


\title{
Self-learning Monte Carlo method with Behler--Parrinello neural networks
}
\author{Yuki Nagai}
\affiliation{CCSE, Japan  Atomic Energy Agency, 178-4-4, Wakashiba, Kashiwa, Chiba, 277-0871, Japan}

\affiliation{
Mathematical Science Team, RIKEN Center for Advanced Intelligence Project (AIP), 1-4-1 Nihonbashi, Chuo-ku, Tokyo 103-0027, Japan
}

\author{Masahiko Okumura}
\affiliation{CCSE, Japan  Atomic Energy Agency, 178-4-4, Wakashiba, Kashiwa, Chiba, 277-0871, Japan}

\author{Akinori Tanaka}
\affiliation{
Mathematical Science Team, RIKEN Center for Advanced Intelligence Project (AIP), 1-4-1 Nihonbashi, Chuo-ku, Tokyo 103-0027, Japan
}
\affiliation{
Department of Mathematics, Faculty of Science and Technology, Keio University, 3-14-1 Hiyoshi, Kouhoku-ku, Yokohama 223-8522, Japan
}
\affiliation{
interdisciplinary Theoretical \& Mathematical Sciences Program (iTHEMS) RIKEN 2-1, Hirosawa, Wako, Saitama 351-0198, Japan
}

\date{\today}
             
\begin{abstract}
We propose a general way to construct an effective Hamiltonian in the Self-learning Monte Carlo method (SLMC), which speeds up 
Monte Carlo simulations by training an effective model to propose uncorrelated configurations in the Markov chain.
Its applications are, however, limited. This is because it is not obvious  to find the explicit form of the effective Hamiltonians. 
Particularly, it is difficult to make trainable effective Hamiltonians including many-body interactions. 
In order to overcome this critical difficulty, we introduce the Behler--Parrinello neural networks (BPNNs) as ``effective Hamiltonian'' without any prior knowledge, which is used to construct the potential-energy surfaces in interacting many particle systems for molecular dynamics.
We combine SLMC with BPNN by focusing on a divisibility of Hamiltonian and propose how to construct the element-wise configurations. 
We apply it to quantum impurity models. 
We observed significant improvement of the acceptance ratio from 0.01 (the effective Hamiltonian with the explicit form) to 0.76 (BPNN). This drastic improvement implies that the BPNN effective Hamiltonian includes many body interaction, which is omitted in the effective Hamiltonian with the explicit forms. The BPNNs make SLMC more promising.

\end{abstract}

\maketitle

\section{Introduction}
Quantum Monte Carlo (QMC) is one of the unbiased numerical methods for studying quantum many-body systems\cite{Blakendecler,Hirsch,HirschPRL,White}. 
The developments of the continuous-time QMC have made great successes in strongly correlated electron systems\cite{Rubtsov,Werner,GullRMP,GullEPL}. 
In this algorithm, the partition function is expanded in a power series of the perturbation terms. 
Both the number and position of perturbation terms on the imaginary-time interval change constantly during the simulation. 
To compute the weight of each configuration, the continuous-time QMC methods require integrating out the degree of freedom of the fermions, which is very time consuming. 

The self-learning Monte Carlo (SLMC)\cite{Liu,LiuShen,Xu,Nagai} was recently introduced as a general method, which speeds up the MC simulation by designing and training a model to propose efficient global updates. 
There are many kinds of applications, such as classical statistical mechanics models\cite{Liu,HuangWang}, classical spin-fermion models\cite{LiuShen}, determinant QMC\cite{Xu,ZHLiu,Chen}, continuous-time QMC\cite{HuangYang,Nagai}, and hybrid MC in high-energy physics\cite{Tanaka}.
The SLMC is one of the successes in machine learning techniques in physics \cite{Carrasquilla,Carleo,Hu,Deng,TanakaTomiya,Fujita,YiZhang,JChen,YHuang,ZiCai,SJohann,ENieuwenberg,Zdeborova,Shiba,Mano}. 
The philosophy behind SLMC is ``first learn, then earn''. 
In the learning stage, we perform trial simulations to generate a large set of configurations and their weights. 
These configurations and weights are then used to train an effective model $H_{\rm eff}$, whose Boltzmann weight $e^{- \beta H_{\rm eff}}$ fits the probability distribution of the original problem. 
Next, in the actual simulation, $H_{\rm eff}$ is used as a guide to propose highly efficient global moves in configuration space. 

A good effective model makes simulations with the SLMC more efficient. 
The efficient effective model is usually invented based on the human understanding of the original system\cite{Liu,LiuShen,Xu,Nagai,ZHLiu}. 
However, it is not easy to construct effective models including many-body interactions, since we do not know systematic procedure applicable in arbitrary systems.
Only for the Hirsch-Fye QMC method in a quantum impurity model, the convolutional deep neural network (CNN) was proposed  to construct the effective Hamiltonian without explicit forms\cite{HuitaoNN}.

The machine learning including artificial neural networks has been used for about twenty years in the field of the molecular dynamics (MD) to construct the potential-energy surfaces (PESs) providing inter-atom forces with accuracy and computational complexity respectively comparable to quantum and classical mechanical calculations.
In the method, the neural network is trained using a large data set
consisting of pairs of an atom configuration with continuous position
index and corresponding total energy in some systems given by quantum
mechanical calculation (e.g., the density functional theory)\cite{Behler2007}. 
We point out that the neural network PESs can be considered as a general scheme to construct effective Hamiltonians of systems consisting of interacting particles with continuous indices. The wide applicability of this method allows us to apply it to complex problems like the imaginary-time MC calculation of electrons in a solid, which has both discrete and continuous coordinates corresponding to positions on a lattice and imaginary-time, respectively.

 \begin{figure*}[t]
\begin{center}
     \begin{tabular}{p{1.5 \columnwidth}} 
      \resizebox{1.5 \columnwidth}{!}{\includegraphics{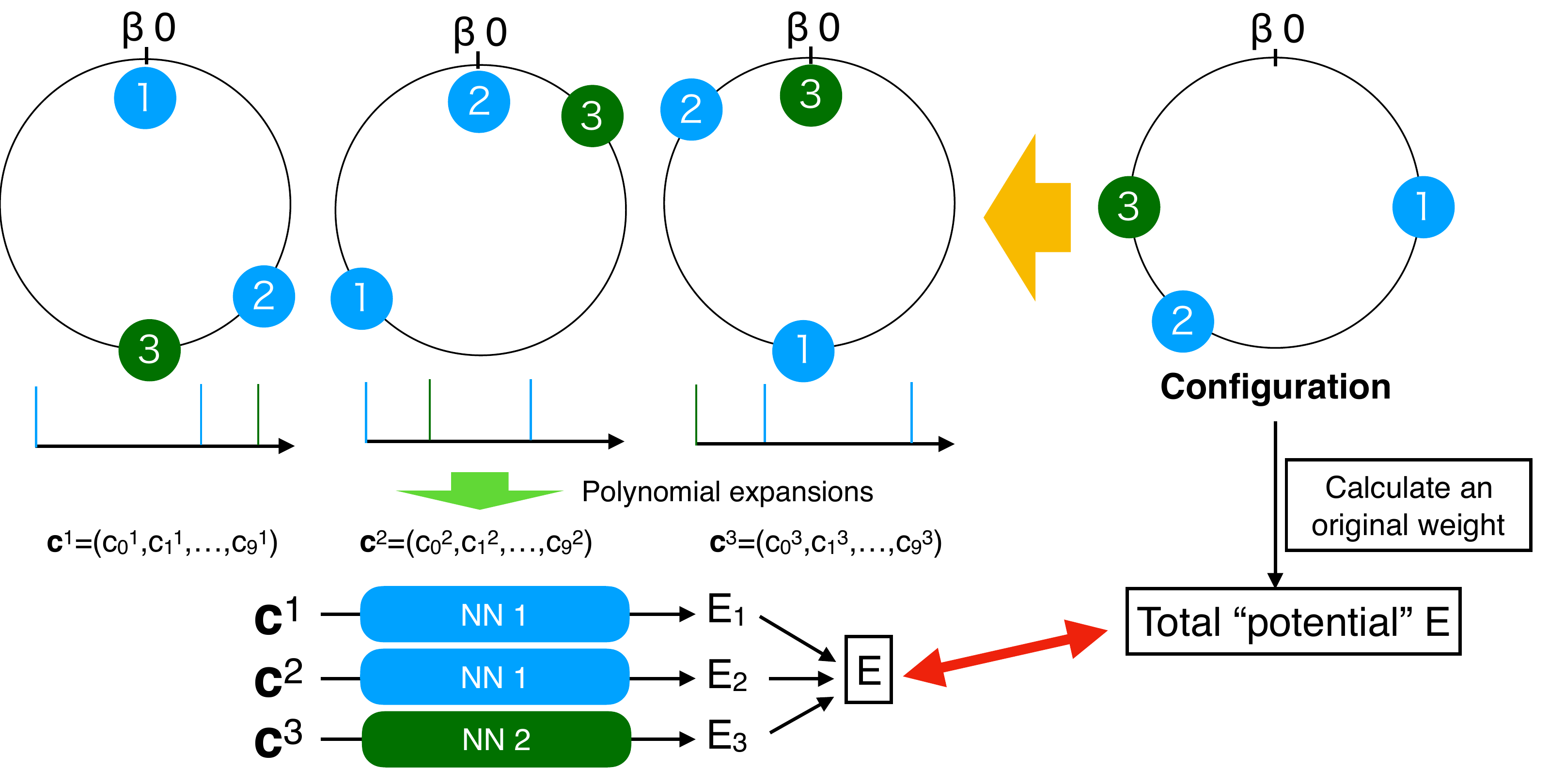}} 
    \end{tabular}
\end{center}
\caption{
\label{fig:fig1}
(Color online) 
Schematic figures of Behler-Parrinello neural networks for the
self-learning continuous-time interaction-expansion quantum Monte Carlo
method. 
The configuration with $N$ vertices on the imaginary-time axis ${\cal C}
= \{ \tau_1,\tau_2,\cdots,\tau_N\}$ is mapped to the set of the
``element-wize'' configurations ${\cal C}^{\rm ele}({\tau_i -\tau_j})$ around a vertex
$j$: ${\cal C} \rightarrow \{ {\cal C}^{\rm ele}({\tau_1 -\tau_j}),{\cal
C}^{\rm ele}({\tau_2 -\tau_j}),\cdots,{\cal C}^{\rm ele}({\tau_N
-\tau_j}) \}$. 
The element-wise configuration is expressed as a distribution function with $\delta$ functions. 
The total potential $E$ of the original calculation is expressed as $E = \log w({\cal C})/(-\beta)$. 
$E$ obtained by the neural networks is expressed as $E = \log w^{\rm eff}({\cal C})/(-\beta) =H_{\rm eff}({\cal C})$. 
The partial energy $E_i$ is defined as $E_i = h_{\rm eff}^{\alpha_i}(\Vec{c}^j)/(-\beta)$. 
The color of the circle denotes a kind of a vertex. Same neural networks are used if the kind of a vertex is same. 
 }
\end{figure*}

In this paper, we propose a method to construct the effective Hamiltonians with Behler--Parrinello neural networks (BPNN)\cite{Behler2007}, which is one of most succeeded methods in the field of the molecular dynamics with machine learning.
We regard the the configuration and effective Hamiltonian in SLMC as the positions of the atoms and the PESs in the MD, respectively. 
We use the recently proposed method \cite{Artrith2017} in the MD field to map continuous coordinates (atom positions) onto discrete variables (inputs of BPNN) , whose advantage is availability of systematic improvement of the mapping accuracy.
As a concrete example, we demonstrate self-learning continuous-time interaction-expansion (CTINT) QMC with BPNNs on quantum impurity models. 
We implement the simplest neural networks and test their performances. 
We also develop the fast updates which is applicable even with deep neural networks to reduce the computational cost significantly in the SLMC simulation. 

The paper is organized as follows. 
The self-learning Monte Carlo method is introduced in Sec. II, 
the effective model in the SLMC is discussed in Sec. III., 
the SLMC and BPNN is combined in Sec.IV, 
an application of the SLMC to quantum impurity models is demonstrated in Sec. V, the discussion is given in Sec. VI, and 
the conclusion is given in Sec.VII. 
In the appendix A, we briefly describe the Markov chain Monte Carlo method. 
The Machine-learning technique in molecular dynamics and Behler--Parrinello neural networks are introduced in the appendix B.
In the appendix C, the technical details of the batch-atom normalization used in the self-learning CTINT.

\section{Self-learning Monte Carlo Method}
\subsection{Metropolis-Hastings algorithm}
The Markov chain Monte Carlo method (MCMC) is a powerful method for an integration in high-dimensional space. 
In physics, the partition functions and physical exception values are often calculated by the MCMC. 
Some difficulties of the MCMC method in physics are described in Appendix \ref{app:markov}.

In the Monte Carlo method, we have to generate a configuration ${\cal C}$  with a probability distribution $w({\cal C})$.
By constructing a Markov chain that has the desired distribution as its equilibrium distribution, 
we can obtain a sample of the desired distribution by observing the chain after a number of steps. 
To construct the Markov chain, we introduce the condition of the detailed balance expressed as 
\begin{align}
w({\cal C}) P({\cal C}' |{\cal C}) &= w({\cal C}') P({\cal C} |{\cal C}') . \label{eq:detail}
\end{align}
Here, $P({\cal C}' |{\cal C})$ is the probability of transitioning from another configuration ${\cal C}$ to a configuration ${\cal C}'$. 
The Metropolis-Hastings approach is to separate the transition in two sub-steps: 
\begin{align}
P({\cal C}' |{\cal C}) &= g({\cal C}' |{\cal C})A({\cal C}',{\cal C}), 
\end{align}
where the proposal distribution $g({\cal C}' |{\cal C})$ is the conditional probability of proposing a configuration ${\cal C}'$ when a configuration ${\cal C}$ is given, and 
the acceptance ratio $A({\cal C}',{\cal C})$ is the probability to accept the proposed configuration ${\cal C}'$. 
By inserting the above expression into Eq.~(\ref{eq:detail}), we have 
\begin{align}
\frac{A({\cal C}',{\cal C})}{A({\cal C},{\cal C}')} &= \frac{w({\cal C}')}{w({\cal C})} \frac{g({\cal C} |{\cal C}')}{g({\cal C}' |{\cal C})}.
\end{align}
The Markov chain that has the desired distribution $w({\cal C})$ is obtained when the acceptance ratio is given as\cite{Bishop}
\begin{align}
A({\cal C}',{\cal C}) &= {\rm min} \left(1,  \frac{w({\cal C}')}{w({\cal C})} \frac{g({\cal C} |{\cal C}')}{g({\cal C}' |{\cal C})} \right). \label{eq:acc}
\end{align}
Then, we can generate the Markov chain expressed as
\begin{align}
{\cal C}_1 \rightarrow \cdots \rightarrow {\cal C}_i \rightarrow \cdots.
\end{align}

One can design various kinds of the Monte Carlo method based on Eq.~(\ref{eq:acc}) (see Appendix \ref{app:markov}). 
We need an update method from ${\cal C}$ to ${\cal C}'$ with the high acceptance ratio for good efficiency of the MCMC. 
The most simple update method is so-called local update, where the configuration is updated locally. 
One randomly chooses a single site in the current configuration and proposes a new configuration by changing the variable on this site. 
However,  the local update suffers heavily from a critical slowing down when the system is close to phase transitions. 
In such cases, the autocorrelation time within the Markov chain $\tau$ becomes very large.
To overcome the increase of the autocorrelation time for the local update, various kinds of global update method have been developed \cite{Swendsen,Wolff,Prokofev,EvertzPRL,Evertz}. 
In all these global update methods, variables on a large number of sites are simultaneously changed in a single Monte Carlo update.  
For a given generic model, it is hard to design an efficient global update method. 

We note that the hybrid Monte Carlo (HMC) method is known as the one of the good global update methods, which is widely used in the lattice quantum chromodynamics (QCD) (see, Appendix \ref{app:hmc}). 
Recently, the HMC method is revisited in the condensed matter physics to treat strongly correlated electron systems\cite{Beyl}, 
since the HMC method is general to obtain uncorrelated configurations. 
Although the HMC method might be suitable for reducing the autocorrelation time, its computational cost is not small.

\subsection{Basic concept of the SLMC}
In MCMC, we can design the proposal probability $g({\cal C} |{\cal C}')$. 
If the ratio of the probability $g({\cal C} |{\cal C}')/g({\cal C'} |{\cal C}) = w({\cal C})/w({\cal C}' )$, the new configuration ${\cal C}'$ 
is always accepted (i.e. $A({\cal C}',{\cal C}) = 1$).
To design the proposal probability, we use the another Markov chain with the probability $w_{\rm prop}({\cal C})$.
We consider that the configuration ${\cal C}'$ is obtained by the random walk from ${\cal C}$ on this proposal Markov chain. 
Its detailed balance condition is given as 
\begin{align}
w_{\rm prop}({\cal C}) P_{\rm prop}({\cal C}' |{\cal C}) &= w_{\rm prop}({\cal C}') P_{\rm prop}({\cal C} |{\cal C}'). \label{eq:detailprop}
\end{align}
This proposal probability $P_{\rm prop}({\cal C}' |{\cal C})$ can be regarded as the conditional probability $g({\cal C}' |{\cal C})$ on the original Markov chain, which proposes a configuration ${\cal C}'$ from given ${\cal C}$. 
Thus, with the use of the relation:
\begin{align}
\frac{g({\cal C} |{\cal C}')}{g({\cal C'} |{\cal C})}  &= \frac{w_{\rm prop}({\cal C}) }{w_{\rm prop}({\cal C}') }, 
\end{align}
the acceptance ratio in the SLMC is given as 
\begin{align}
A({\cal C}',{\cal C}) &= {\rm min} \left(1,  \frac{w({\cal C}')}{w({\cal C})} \frac{w_{\rm prop}({\cal C}) }{w_{\rm prop}({\cal C}') }\right). \label{eq:accslmc}
\end{align}
If we can design the proposal Markov chain whose probability is equal to that of the original Markov chain $w_{\rm prop}({\cal C}) = w({\cal C})$, 
the proposed configuration ${\cal C}'$ is always accepted. 
The average acceptance rate $\langle A \rangle$ can be estimated by 
$\langle A \rangle = \exp \left[ -\sqrt{\rm MSE} \right]$  
with the mean squared error ${\rm MSE} = (1/n) \sum_{i} (\ln w_{\rm prop}({\cal C}_i) -\ln w({\cal C}_i))^{2}$. \cite{HuitaoNN} 
Here, $n$ is the number of the measurements in Monte Carlo simulations. 

The difference between several update methods is shown in Table \ref{table:1}.
HMC and SLMC are the global updates where the configurations are changed globally. 
In the SLMC, the configuration is proposed by the proposal Markov chain. 
We have to find a good proposal Markov chain whose probability $w_{\rm prop}({\cal C})$ is similar to original one $w({\cal C})$. 

\begin{table}[t]
\caption{Difference between three update methods}
\label{table:1}
\begin{ruledtabular}
\begin{tabular}{cccccc}
 &
Hand-designed &
HMC & 
SLMC  \\
Propose method & by hand &
MD  &
Markov chain  \\
$g({\cal C} |{\cal C}')/g({\cal C'} |{\cal C})$ & usually 1 & 
1 &
$w_{\rm eff}({\cal C})/w_{\rm eff}({\cal C}' )$ 
\end{tabular}
\end{ruledtabular} 
\end{table}

To design the proposal Markov chain, we have to construct an effective model. 
We introduce an effective Hamiltonian $H_{\rm eff} ({\cal C})$:
\begin{align}
w_{\rm prop}({\cal C})  = w^{\rm eff}({\cal C}) &= \exp \left[ - \beta H_{\rm eff}({\cal C}) \right]. \label{eq:wc}
\end{align}
This model $H_{\rm eff} ({\cal C})$ can be constructed by a supervised machine-learning technique\cite{Liu}. 
In the learning stage, trial simulations are performed to generate a large set of configurations and their weights. 
These data are then used to train an effective model $H_{\rm eff} ({\cal C})$, whose weight fits the probability distribution of the original problem $w({\cal C})$. 
Next, in the actual simulation, $H_{\rm eff} ({\cal C})$ is used as a guide to propose highly efficient global moves in configuration space. 
It is important to obtain good effective models in the SLMC simulations.  
In the previous study\cite{Nagai}, we have successfully obtained the form of the effective Hamiltonian with two-body interactions in the continuous-time auxiliary-field QMC (CTAUX) for the Anderson impurity model. 
In the CTINT simulations, 
Huang {\it et al.} have produced the classical Hamiltonian with two- and three- body interactions to reproduce the weights\cite{HuangYang}. 
However, it seems hard to construct the effective Hamiltonian in other systems or other methods. 

\subsection{Calculation cost of the SLMC: reduction of the autocorrelation time}\label{sec:cost}
\begin{figure}[t]
    \begin{center}
         \begin{tabular}{p{1 \columnwidth}} 
          \resizebox{1 \columnwidth}{!}{\includegraphics{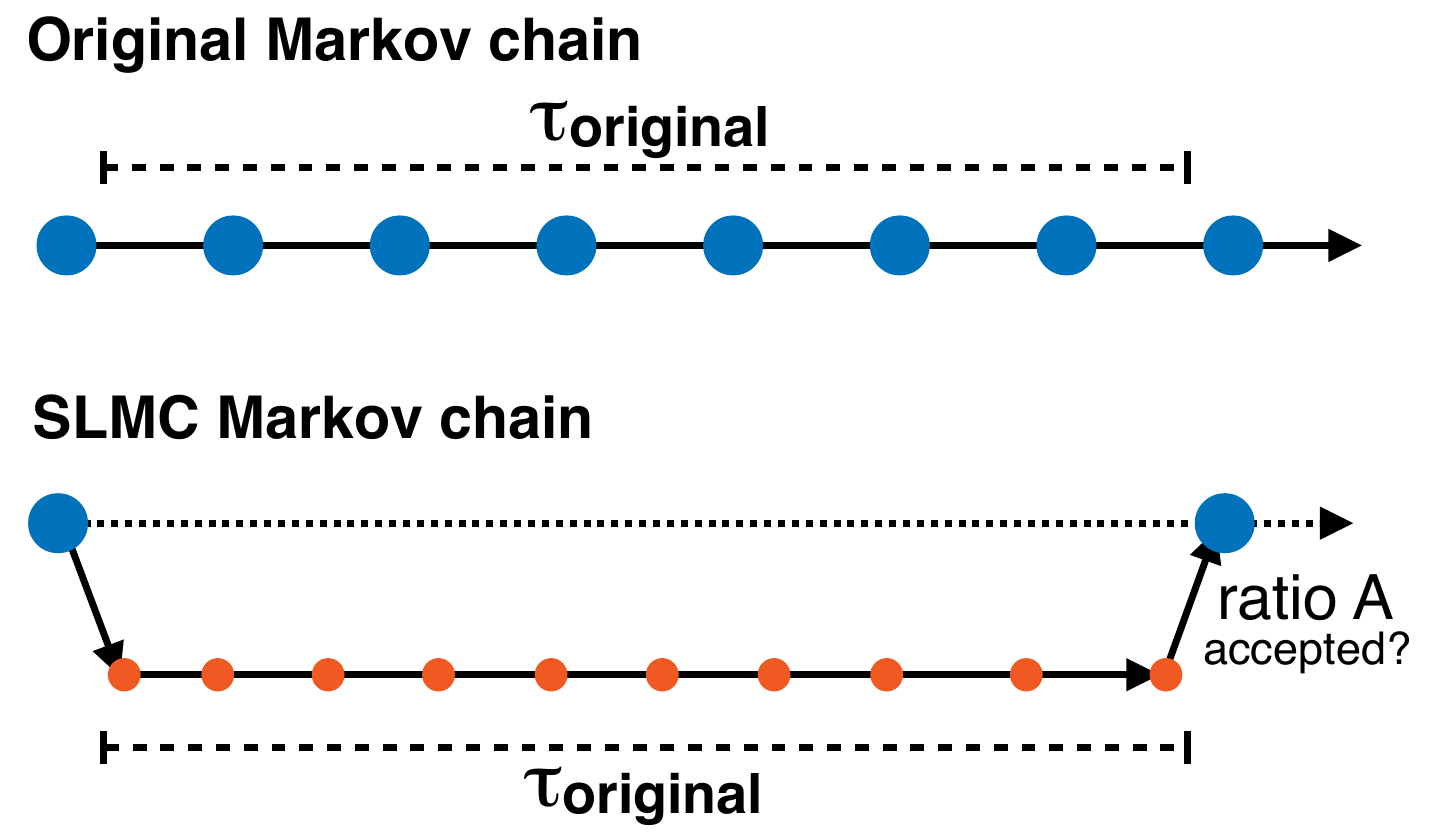}} 
        \end{tabular}
    \end{center}
    \caption{
    \label{fig:slmc}
    (Color online) 
    Schematic figure for the Markov chains of the original Monte Carlo and SLMC. 
    The SLMC can bypass links of the computational costly weight estimations. 
    Large blue (small red) circles denote the computational complexity $u_{\rm original}$ ($u_{\rm SLMC}$) for the calculating the weight of the original model $w({\cal C})$ ($w({\cal C})$). 
    $\tau_{\rm original}$ is the autocorrelation time of the original MC simulation. 
    Since a complexity of a matrix determinant for Fermion systems is quite large, $u_{\rm SLMC} < u_{\rm original}$ is satisfied even if the neural networks are adopted.
    If the average acceptance ratio $\langle A \rangle$ is not too small, the autocorelation time calculated as the number of the large blue circles is much shorter than that of the original simulation. 
     }
    \end{figure}

A computational cost of MCMC simulations is usually estimated by autocorrelation time. 
However, autocorrelation time depends on a kind of the correlation function which we focus on. 
As shown in Fig.~\ref{fig:slmc}, the SLMC can bypass links of the computational costly weight estimations. 
Thus, in SLMC, autocorrelation time for all correlation function is shorter than that in the original simulation. 
We should note that, although the elapsed time comparison in acucual calculations might be regarded as the other estimation of the computational cost, this estimation can not be a fair estimation in the case of the SLMC, since the elapsed time of the original MC and SLMC both depends on the computer architecture and coding techniques of optimazation.

We propose the general estimation of the computational cost of the SLMC as follows.
We define $u_{\rm original}$  as the computational complexity for calculating the weight of the original model $w({\cal C})$.
We consider the autocorrelation time of the original MC simulation $\tau_{\rm original}$. 
The total calculation cost of the original simulation is written as 
\begin{align}
U_{\rm original} \propto \tau_{\rm original } u_{\rm original}. 
\end{align}
On the other hand, the total calculation cost of the SLMC is written as 
\begin{align}
U_{\rm SLMC} &\propto \frac{1}{\langle A \rangle}(N_{\rm steps}^{\rm prop} u_{\rm SLMC} + u_{\rm original}), \\
&= \left( 
\frac{N_{\rm steps}^{\rm prop} }{\langle A \rangle} \frac{u_{\rm SLMC}}{u_{\rm original}} + \frac{1}{\langle A \rangle}
\right)
u_{\rm original} \label{eq:autoslmc}
\end{align}
Here, $\langle A \rangle$ is the average acceptance ratio and $N_{\rm steps}^{\rm prop}$ is the length of the proposal Markov chain. 
We define $u_{\rm SLMC}$ as the computational complexity for calculating the weight of the original model $w({\cal C})$.
The ratio of the computational cost is given as 
\begin{align}
\frac{U_{\rm SLMC} }{U_{\rm original}} \propto \frac{1}{\langle A \rangle} \left(
\frac{N_{\rm steps}^{\rm prop} }{\tau_{\rm original } } \frac{u_{\rm SLMC}}{u_{\rm original}} + \frac{1}{\tau_{\rm original }}
 \right).
\end{align}
We note that $N_{\rm steps}^{\rm prop} = \tau_{\rm original }$ if the proposal Markov chain uses the same update method of the original simulation.
If the effective Hamiltonian gives a perfect weight ({\it i.e.}, $w^{\rm eff}({\cal C})  = w({\cal C})$), the average acceptance ratio is always one: $\langle A \rangle = 1$. 
In this case, if the length of the proposal Markov chain $N_{\rm steps}^{\rm prop}$ is longer than $\tau_{\rm original}$, there is no correlation between the previous and proposed configurations ${\cal C}$ and ${\cal C}'$. 
The calculation cost depends on the ratio of $u_{\rm SLMC}/u_{\rm original}$. 
Thus, the SLMC is faster than the original simulation when the computational cost for effective weight $w^{\rm eff}({\cal C})$ is smaller than that for  the original weight $w({\cal C})$ and the average acceptance ratio $\langle A \rangle$ is not too small. 

With the use of Eq.~(\ref{eq:autoslmc}), the number of the weight calculation $u_{\rm original}$ can be regarded as the definition of the autocorrelation time of the SLMC.
Thus, the autocorrelation time $\tau_{\rm SLMC}$ is expressed as 
\begin{align}
\tau_{\rm SLMC} \sim  \frac{1}{\langle A \rangle}.
\end{align}

\section{Effective models}
The critical problem in SLMC is construction of an effective Hamiltonian $\ln w^{\rm eff}({\cal C})$ which mimics the original one $\ln w({\cal C})$.  
In this section, we introduce the Behler-Parrinello neural networks as a general method to construct the effective Hamiltonian.

\subsection{Effective Hamiltonians}
We discuss the most simplest effective Hamiltonian is one with only two-body interaction terms.
It can be represented as 
\begin{align}
H_{\rm eff} ({\cal C}) = \sum_{ij}^M L({\cal C}_i,{\cal C}_j),
\end{align}
where ${\cal C}_i$ is the $i$-th element of the configuration ${\cal C} = ({\cal C}_1,\cdots,{\cal C}_M)$, and $M$ is the number of the elements. 
The element is usually coordinate of the ``atoms'' in the imaginary-time and/or real space. 
In the previous work, we introduced an explicit form of $L({\cal C}_i,{\cal C}_j)$ which has been known.
But, in general, it is not easy to find explicit forms of effective Hamiltonians even including only two body interactions for arbitrary Hamiltonians.

\subsection{Go beyond the two-body interaction: Behler-Parrinello neural networks}
We introduce the BPNNs as a general method to construct effective Hamiltonians including many-body interaction in the SLMC. 
Before showing detailed formulations, we give brief discussions on the BPNNs and its applicability for SLMC.

In the research field of MD, the machine learning techniques have been used for about twenty years to evaluate effective inter-atom or inter-molecule forces.
The BPNNs method is one of most succeeded methods using them.
See Appendix \ref{app:bpnn} for details of the BPNNs based method.
Originally, the BPNNs are used to obtain forces among atoms without time-consuming calculations of {\it ab initio} MD.
In the method, the forces are obtained as derivatives of a PES with respect to atom positions, which is represented by artificial neural networks trained using data sets obtained by {\it ab initio} MD.
It should be mentioned that the method provides the PES including not only two-body but many-body interactions which are contained in {\it ab initio} MD.

\begin{table}[t]
\caption{Similarity between {\it ab initio} MD and QMC}
\label{table:2}
\begin{ruledtabular}
\begin{tabular}{ccccc}
 &
MD &
QMC  \\
Input & atom-positions & 
configurations \\
Time-consuming & energy &
MC weight \\
Eff. Hamiltonian (2-body)  & BP & SLMC \\
Eff. Hamiltonian (many-body)  & BP & --- \\
\end{tabular}
\end{ruledtabular} 
\end{table}

In order to clarify applicability of the BPNNs for SLMC, let us discuss the similarity between {\it ab initio} MD and QMC.
Correspondences are shown in Table~\ref{table:2}. 
Most time consuming part in the whole calculation is evaluation of energy which depends on the coordinates.
The BP method avoid the bottleneck of the calculations, i.e., DFT calculations, using the PES represented by trained artificial neural networks.
On the other hand, in MC simulations, the Monte Carlo weights are critical quantity.
Calculation of them is most time consuming in the MC simulations.
SLMC is one of promising methods to shorten the calculation time on the part by replacing the Hamiltonian with the effective one.
But, as mentioned above, there were no systematic way to find the effective Hamiltonian especially including many-body interactions (Table \ref{table:2}).
The BP method can give the effective Hamiltonian depending on the configuration in MC, which is completely equivalent with the PES depending on the atom coordinates.

\section{Combine SLMC with BPNN}
\subsection{Divisibility of Hamiltonian}

The discussion in the previous section clarified general correspondence between these two methods.
In this section, we show deeper correspondence between them in fermion systems.

In the BP method, the total energy in the system $E_{\rm tot}^{({\rm S})} \! \left( \mathcal{R}^{({\rm S})} \right)$ is assumed to be divided into ``partial energies'' associated with each atoms in the system, which are determined by ``environment'' around each atoms, i.e., 
\begin{equation}
    E_{\rm tot}^{({\rm S})} \! \left( \mathcal{R}^{({\rm S})} \right) =
    \sum_{i=1}^{N^{({\rm s})}} E_{\rm part}^{({\rm S})} \! \left( \Delta^{({\rm S})}_i
    \right) \, , \label{EtotEpart}
\end{equation}
where
\begin{equation}
\Delta^{({\rm S})}_i = \left\{ {\bm r}_{ij} \, \big| \, j = 1,
\cdots, i - 1, i+1, \cdots , N^{({\rm s})} \right\} \label{delSs}
\end{equation}
is a set of relative coordinates with respect to the $i$-th atom (${\bm r}_{ij} = {\bm r}_j - {\bm r}_i$).
This divisibility of the Hamiltonian is an assumption, but the artificial neural networks are trained to find a set of the partial energies to reproduce the original Hamilton.

On the other hand, in the MC calculations of fermion systems, the primary quantity is the partition function $Z$ expressed as the determinant of the matrix:
\begin{align}
Z = \sum_{\cal C} {\rm det} M({\cal C}) = \sum_{\cal C} w({\cal C}).
\end{align}
The dimension of the matrix depends on variation of the MC methods. 
For example, in continuous-time QMC (e.g. CT-AUX and CT-INT) for an impurity model, the dimension of the matrix $M$ is the number of the vertices on the imaginary-time axis. 
Here, we mention that the logarithm of the weight $\ln w({\cal C})$ can be divided into $N$ parts as 
\begin{align}
\ln w({\cal C}) &=  \ln {\rm det} M({\cal C}), \\
&= \sum_{j=1}^N \left[ \ln M({\cal C}) \right]_{jj} \, . \label{lnwsum}
\end{align}
This equation leads partitioning of an arbitrary effective Hamiltonian into the partial Hamiltonians, i.e.,
\begin{align}
H_{\rm eff}({\cal C}) & = \sum_{j=1}^N h_{\rm
eff}^{\alpha_{j}}(N,\Vec{c}^j) \, . \label{eq:hlocals}
\end{align}
We emphasize that this equation is not an assumption but an exact expression.

There is no rigorous correspondence between them although Eqs.~(\ref{EtotEpart}) and (\ref{lnwsum}) have similar expressions.
In this paper, we show, however, the BP method give appropriate partitioning of the effective Hamiltonian.

\subsection{Effective Hamiltonian in SLMC}
In actual SLMC simulations, 
we adopt the effective Hamiltonian defined as 
\begin{align}
    H_{\rm eff}({\cal C}) &= \frac{1}{N} \sum_{j=1}^N h_{\rm eff}^{\alpha_{j}}(\Vec{c}^j) + f(N), \label{eq:effh}
\end{align}
Here, $f(N)$ is a polynomial function $f(N) = \sum_{k=0}^{n_{\rm max}} f_{k} N^{k}$. 
We assume t $h_{\rm eff}^{\alpha_{j}}(N,\Vec{c}^j) = (1/N) h_{\rm eff}^{\alpha_{j}}(\Vec{c}^j)$. 
This factor $1/N$ is introduced as the normalization, which is appropriate for the CTAUX in the previous paper.
The $h_{\rm eff}^{\alpha_j}(\Vec{c}^j)$ is constructed by neural networks. 
Note that the nonlinear function $h_{\rm eff}^{\alpha_j}(\Vec{c}^j)$ does not depend on $j$ and only depends on $\alpha_j$, since we assume that the same kind of atoms feels same interactions.

\subsection{Construction of the element-wise configurations}
We have to generate element-wise configurations $\Vec{c}^j$ from ${\cal C}$.
Usually, in QMC simulations, the configuration ${\cal C}$ is a set of positions of ``atoms'' (e.g. spins or vertices) on real and/or imaginary-time axes. 
We can make the element-wise configurations $\Vec{c}^j$ which consist of the distances between the atom $j$ and other atoms.  
However, it is not a good representation of the configuration for NNs, 
since the number of elements of the input vector $\Vec{c}^j$ varies depending 
on the number of atoms although the number of inputs of NNs must be fixed.
Actually, in continuous-time QMC simulations, the number of atoms changes during simulations.

One of the method to construct the element-wise configurations on continuous axis is the Chebyshev polynomial expansion method in the field of machine-learning MD simulations\cite{Artrith2017}. 
The element-wise configuration is expressed by coefficients of the basis functions as follows.
For simplicity, we consider a configuration which has vertices on the imaginary-time axis such as a quantum impurity model. 
We map the configuration ${\cal C}$ onto the set of the element-wise
configurations ${\cal C}^{\rm ele}({\tau_i -\tau_j})$ around a vertex
$j$: ${\cal C} \rightarrow \{ {\cal C}^{\rm ele}({\tau_1 -\tau_j}),{\cal
C}^{\rm ele}({\tau_2 -\tau_j}),\cdots,{\cal C}^{\rm ele}({\tau_N
-\tau_j}) \}$. 
The element-wise configuration is expressed by the basis functions. 
We introduce the density distribution functions defined as 
\begin{align}
\rho(\tau,{\cal C}^{\rm ele}({\tau_i -\tau_j})) = \sum_{i=1}^N \delta(\tau - \tau_{ij}),
\end{align}
where $\tau_{ij} =  2|\tau_{i}-\tau_{j}|/\beta -1$ is the distance between atom $i$ and atom $j$. 
This distribution is expanded by the Chebyshev polynomial functions \cite{Artrith2017}: 
\begin{align}
\rho(\tau,{\cal C}^{\rm ele}({\tau_i -\tau_j})) = \sum_m c_m^j \phi_m(\tau), \label{eq:rho}
\end{align}
where 
\begin{align}
c_m^j = \sum_{i=1}^N \phi_m(\tau_{ij}). \label{eq:cc}
\end{align}
Here, $\phi_m(x)$ is the Chebyshev polynomial function $\phi_m(x) = \cos (m \arccos(x))$. 
Thus, we map the element-wise configuration ${\cal C}^{\rm ele}({\tau_i
-\tau_j})$ onto the set of $m_{\rm cut}$ coefficients of the Chebyshev
polynomials ${\cal C}^{\rm ele}({\tau_i -\tau_j}) \rightarrow \{c_0^j,\cdots,c_{m_{\rm cut}-1}^j \}$, as shown in Fig.~\ref{fig:fig1}. 
The element-wise configuration $\Vec{c}^j$ is expressed as $\Vec{c}^j \equiv \{c_0^j,\cdots,c_{m_{\rm cut}-1}^j \}$, which does not depend on the number of atoms.

To introduce a difference of species, we can add another distribution functions. 
For example, if the vertex has a spin index $s_j$ and there is spin-reversal symmetry (i.e. the weight is not changed by flipping all spins), it is better to add ``spin-density'' distribution functions defined as 
\begin{align}
\rho_s(\tau,{\cal C}^{\rm ele}({\tau_i -\tau_j})) = \sum_{i=1}^N s_i s_j \delta(\tau - \tau_{ij}),
\end{align}
With the use of the Chebyshev polynomial expansions, we have 
\begin{align}
\rho_s(\tau,{\cal C}^{\rm ele}({\tau_i -\tau_j})) = \sum_m d_m^j \phi_m(\tau), \label{eq:rhos}
\end{align}
where 
\begin{align}
d_m^j = \sum_{i=1}^N s_i s_j \phi_m(\tau_{ij}).\label{eq:dc}
\end{align}
Then, the element-wise configuration $\Vec{c}^j$ is $\Vec{c}^j \equiv \{c_0^j,\cdots,c_{m_{\rm cut}-1}^j, d_0^j,\cdots,d_{m_{\rm cut}-1}^j \}$. 

\subsection{Relation between the SLMC with BPNN and previous effective model}
In the previous paper about the CTAUX QMC, the configuration ${\cal C}$ consists of vertices with a spin index on the imaginary-time continuous axis \cite{Nagai}.
We show that the effective Hamiltonian in this previous paper can be regarded as Eq.~(\ref{eq:effh}) with linear functions $h_{\rm eff}^{\alpha_j}(\Vec{c}^j)$ without hidden layers. 
If the function $h_{\rm eff}^{\alpha_j}(\Vec{c}^j)$  is linear, the effective Hamiltonian is expressed as 
\begin{align}
    H_{\rm eff}({\cal C}) &= \frac{1}{N} \sum_{j=1}^N W^T \Vec{c}^j + b+ f(N). 
\end{align}
Here, $W$ is a $M$-dimensional vector and $M$ is the number of the input elements.
With the use of the element-wise configuration $\Vec{c}^j$ is $\Vec{c}^j \equiv \{c_0^j,\cdots,c_{m_{\rm cut}-1}^j, d_0^j,\cdots,d_{m_{\rm cut}-1}^j \}$, 
this effective Hamiltonian is rewritten as 
\begin{align}
    H_{\rm eff}({\cal C}) &= \frac{1}{N} \sum_{j=1}^N  \sum_{m=1}^{m_{\rm cut}} W_{l}^c c_{l-1}^j 
    + \frac{1}{N} \sum_{j=1}^N \sum_{m=1}^{m_{\rm cut}} W_{l}^d d_{l-1}^j
    +f(N). 
\end{align}
Here, the coefficient $b$ is included in $f(N)$. 
With the use of Eqs.~(\ref{eq:cc}) and (\ref{eq:dc}), we obtain 
\begin{align}
    H_{\rm eff}({\cal C}) &= \frac{1}{N} \sum_{i,j=1}^N  L(\tau_i -\tau_j) + \frac{1}{N}\sum_{i,j=1}^N  J(\tau_i -\tau_j) s_i s_j +f(N), \label{eq:linear}
\end{align}
where 
\begin{align}
L(\tau_i -\tau_j)  &\equiv \sum_{m=1}^{m_{\rm cut}} W_{l}^c \phi_m(\tau_{ij}), \\
J(\tau_i -\tau_j) &\equiv \sum_{m=1}^{m_{\rm cut}} W_{l}^d \phi_m(\tau_{ij}).
\end{align}
This effective Hamiltonian is completely equivalent to that used in the previous paper\cite{Nagai}.

\subsection{Many body interactions with neural networks}
\begin{figure}[t]
\begin{center}
     \begin{tabular}{p{0.7 \columnwidth}} 
      \resizebox{0.7 \columnwidth}{!}{\includegraphics{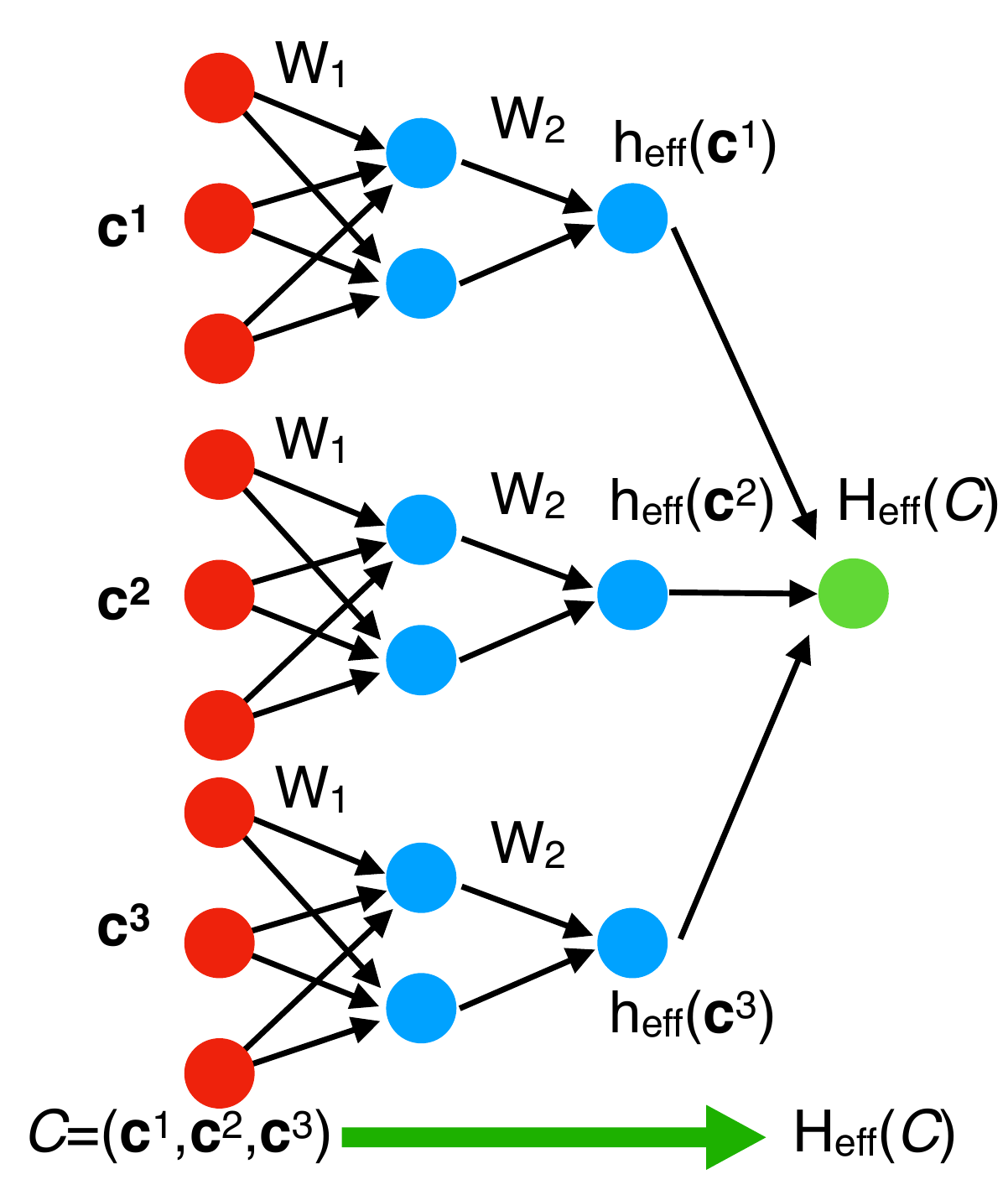}} 
    \end{tabular}
\end{center}
\caption{
\label{fig:fig2}
(Color online) 
Schematic figures of input and output data in Behler-Parrinello neural networks. 
The weight is expressed as a sum of the element-wise effective Hamiltonians. 
If we neglect hidden layers, the effective Hamiltonian is constructed by linear combinations. 
 }
\end{figure}

The effective Hamiltonian with linear functions $h_{\rm eff}^{\alpha_j}(\Vec{c}^j)$ includes only two-body interactions as shown in Eq.~(\ref{eq:linear}).
To include the many-body interactions, we can add hidden layers. 
 For example, in the case of the neural networks with one hidden layer with $N_u$ units, the effective element-wise Hamiltonian is expressed as 
\begin{align}
    h_{\rm eff}^{\alpha_{j}}(\Vec{c}^j) =  \hat{W}_{2}^{\alpha_{j}} F \left( \hat{W}_{1}^{\alpha_{j}} \Vec{c}^j+
\Vec{b}_{1}^{\alpha_{j}}
 \right)+b_{2}^{\alpha_{j}},
\end{align}
with the activation function $[F(\Vec{x})]_{i} = F([\Vec{x}]_{i})$. 
Here, $\hat{W}_{1}^{\alpha_{j}}$ is a $N_{u} \times M$ matrix, $\hat{W}_{2}^{\alpha_{j}}$ is a $1 \times N_{u}$ matrix, 
$\Vec{b}_{1}^{\alpha_{j}}$ is a $N_{u}$-dimensional bias vector,  and $b_{2}^{\alpha_{j}}$ is a bias. 
The activation function $F(x)$ makes the effective Hamiltonian nonlinear, which can represent many-body interactions. 
The schematic figure of the effective Hamiltonian is shown in Fig.~\ref{fig:fig2}. 
We use a common weight matrix $W$ to same atomic species.

\section{Demonstrations on quantum impurity models}
\subsection{Continuous-time QMC for fermions}
In continuous-time QMC simulations, by splitting the Hamiltonian into non-perturbative and perturbative parts $H = H_{0} + H_{1}$, 
the partition function is expanded as 
\begin{align}
Z &= {\rm Tr} \: \left[e^{- \beta H_{0}} T_{\tau} e^{- \int_{0}^{\beta} H_{1}(\tau) d\tau} \right], \\
&=  \sum_{N=0}^{\infty} \frac{(-1)^{N}}{N!} \int_{0}^{\beta} d\tau_{1} \cdots \int_{0}^{\beta} d\tau_{N} {\rm Tr} \left[T_{\tau} e^{-\beta H_{0}} H_{1}(\tau_{N}) \right. \nonumber \\
&\quad {} \times \left. H_{1}(\tau_{N-1}) \cdots H_{1}(\tau_{1}) \right], \\
&= \sum_{{\cal C}} W({\cal C}),
\end{align}
where $T_{\tau}$ is the imaginary-time-ordering operator. 
Here, the configuration ${\cal C}$ has $N$ vertices on the imaginary-time axis\cite{GullRMP}. 
The number of vertices $N$ changes during simulation.

\subsection{Continuous-Time interaction expansion quantum Monte Carlo in impurity models}
We demonstrate self-learning CTINT with BPNNs of the impurity model.
The CTINT method is a good example to demonstrate the SLMC with BPNN, 
since Huang {\it et al.} claimed that the three body interactions are necessary to construct good effective model in CTINT method.\cite{HuangYang}

We consider the Hamiltonian of the single impurity Anderson model, which is written as the combination of the free fermion part and the interaction part\cite{GullRMP,Nagai}, 
\begin{align}
H &= -\mu \sum_{\sigma} n_{\sigma} + \sum_{\sigma,p}(V c_{\sigma}^{\dagger} a_{p,\sigma}+{\rm H. c.}) \nonumber \\
  &\quad {} + \sum_{\sigma,p} \epsilon_{p} a_{p,\sigma}^{\dagger} a_{p,\sigma} + U n_{\up} n_{\down},
\end{align}
where $\sigma = \up, \down$, $c_{\sigma}^{\dagger}$ and $a_{p,\sigma}^{\dagger}$ are the fermion creation operators for an impurity electron with spin $\sigma$, and that for a 
bath electron with spin $\sigma$ and momentum $p$, respectively. $n_{\sigma}$ is the impurity electron number operator. 
We consider a bath with a semicircular density of states $\rho_0(\epsilon) = [2/(\pi D)\sqrt{1-(\epsilon/D)^2}]$ and set the half bandwidth $D = 1$ as the energy unit. 
In the CTINT algorithm, we rewrite the interaction part expressed as 
\begin{align}
H_{1} = \frac{U}{2} \sum_{s=\pm 1} \left( n_{\up} - \frac{\rho}{2} - s \delta \right) \left(n_{\down} -\frac{\rho}{2} + s \delta \right).
\end{align}
Here, we introduce additional Ising variable $s$ and parameter $\delta$, and $\rho$ corresponds to the average electron density\cite{Assaad2007,Assaad,Luitz}. 
We consider the half filling ($\rho = 1$). 
The non-perturbative part is 
$H_0 = \left(-\mu+\frac{U}{2}\rho \right) \sum_{\sigma} n_{\sigma} + \sum_{\sigma,p}(V c_{\sigma}^{\dagger} a_{p,\sigma}+{\rm H. c.}) 
+ \sum_{\sigma,p} \epsilon_{p} a_{p,\sigma}^{\dagger} a_{p,\sigma}$. 
The partition function is expanded as 
\begin{align}
 \frac{Z}{Z_{0}} &= \sum_{{\cal C}} W({\cal C}) = \sum_{\cal C} 
 \left(\frac{-U}{2} \right)^{N}  \frac{1}{N!} \prod_{\sigma} {\rm det} M_{\sigma}({\cal C})
 .
 \end{align}
Here,  the $N \times N$ matrix $M_{\sigma}({\cal C})$ is defined by $[M_{\sigma}({\cal C})]_{ij} \equiv g_{0}(\tau_{i}-\tau_{j}) - \alpha_{\sigma}(s_{n}) \delta_{ij}$ with  $\alpha_{\sigma}(s) \equiv \rho/2+\sigma s \delta$. 
$g_{0}(\tau_{i}-\tau_{j})$ is the free fermion Green's function at the impurity site.
We implement the CTINT with the Julia language 0.6.2.
We train the neural networks with one hidden layer as shown in Fig.~\ref{fig:fig2} with 50000 training data, which is done by the TensorFlow 1.4,  one of the deep learning frameworks, in Python 3.6.5.  
The sigmoid function $F(x) = 1/(1+\exp(-x))$ is used as the activation function. 
We develop the batch-atom normalization, variant of the batch normalization \cite{ioffe2015batch} which is one of the modern techniques accelerating training procedures for neural network by normalizing along batch index.
In the batch-atom normalization, 
we normalize both batch and atomic index $j$ (See, Appendix A). 
The input vector is a $M=2m_{\rm cut}$ dimensional vector: $\Vec{c}^j = (c_0^j,\cdots,c_{m_{\rm cut}-1}^j,d_0^j,\cdots,d_{m_{\rm cut}-1} )^T$.
The total number of parameters in neural networks with one hidden layer with $N_u$ units is $M N_u + N_u + 4 N_u+ N_u + n_{\rm max}+1 = M N_u + 6N_u +  n_{\rm max}+1$. 
We usually set the Chebyshev expansion cutoff $m_{\rm cut} = 10$, which determines the resolution of the input vector on the imaginary-time axis. 

\begin{figure}[t]
\begin{center}
     \begin{tabular}{p{1 \columnwidth}} 
      \resizebox{1 \columnwidth}{!}{\includegraphics{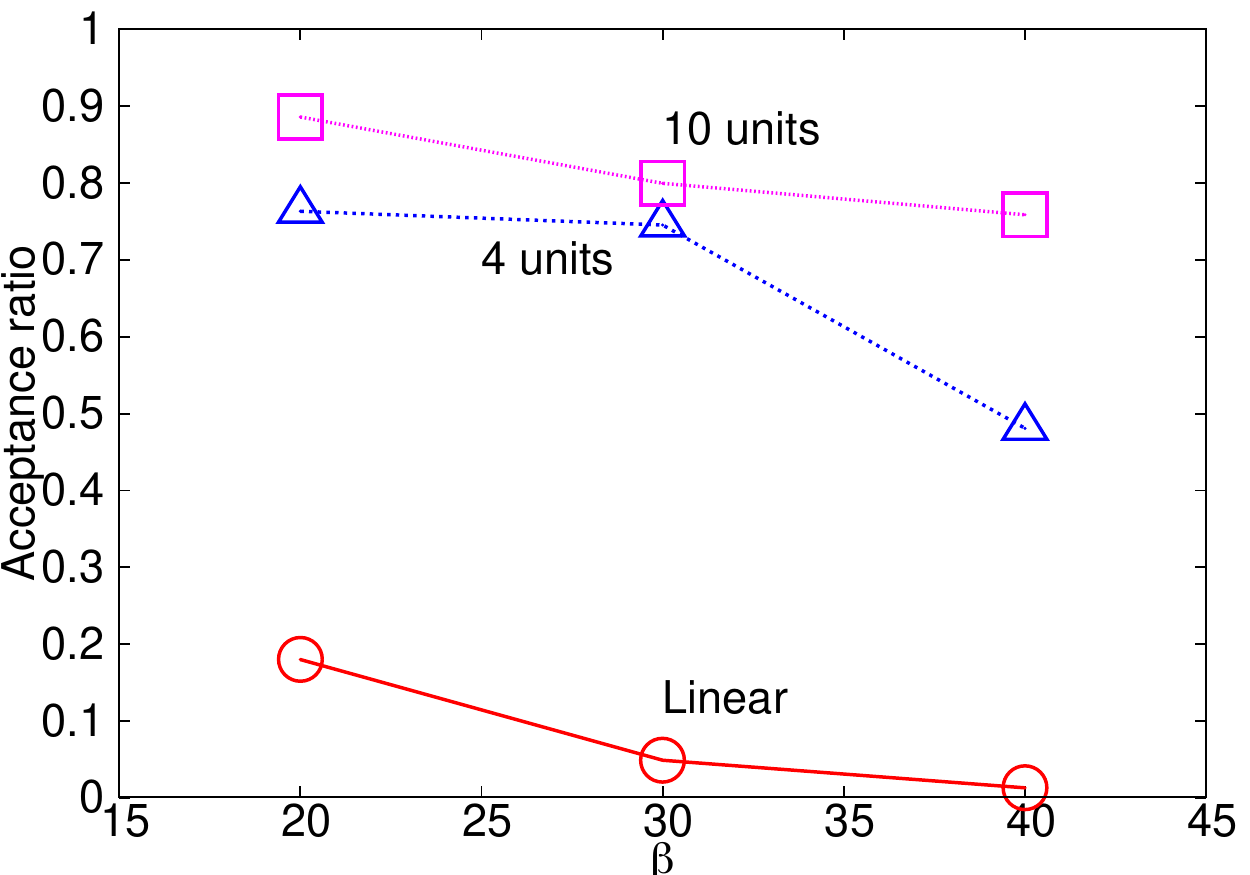}} 
    \end{tabular}
\end{center}
\caption{
\label{fig:fig3}
(Color online) 
Inverse-temperature dependence of the acceptance ratio in the self-learning continuous-time interaction expansion quantum Monte Carlo simulation in the Anderson impurity model. We consider $U=3D$, $\delta = 0.5$, $V = 1D$ and 500 SLMC steps. 
 }
\end{figure}

Figure \ref{fig:fig3} shows the inverse-temperature dependence of the acceptance ratio of the SLMC. 
We consider $U = 3D$, $\delta = 0.5$, $V=1D$, $m_{\rm cut} = 10$, and $n_{\rm max}=3$. 
We set the number of the SLMC steps $n$ is $n=500$.
We also consider the linear SLMC, which is equivalent to the previous effective Hamiltonian\cite{Nagai}.
In the case with $\beta = 40$, the acceptance ratio with $10$ units ($N_u = 10$) is around 0.8, while that of the linear SLMC is less than 0.02. 
The results indicates that the BPNNs can systematically improve the effective Hamiltonian with increasing the number of units.  

\begin{figure}[t]
\begin{center}
     \begin{tabular}{p{1 \columnwidth}} 
      \resizebox{1 \columnwidth}{!}{\includegraphics{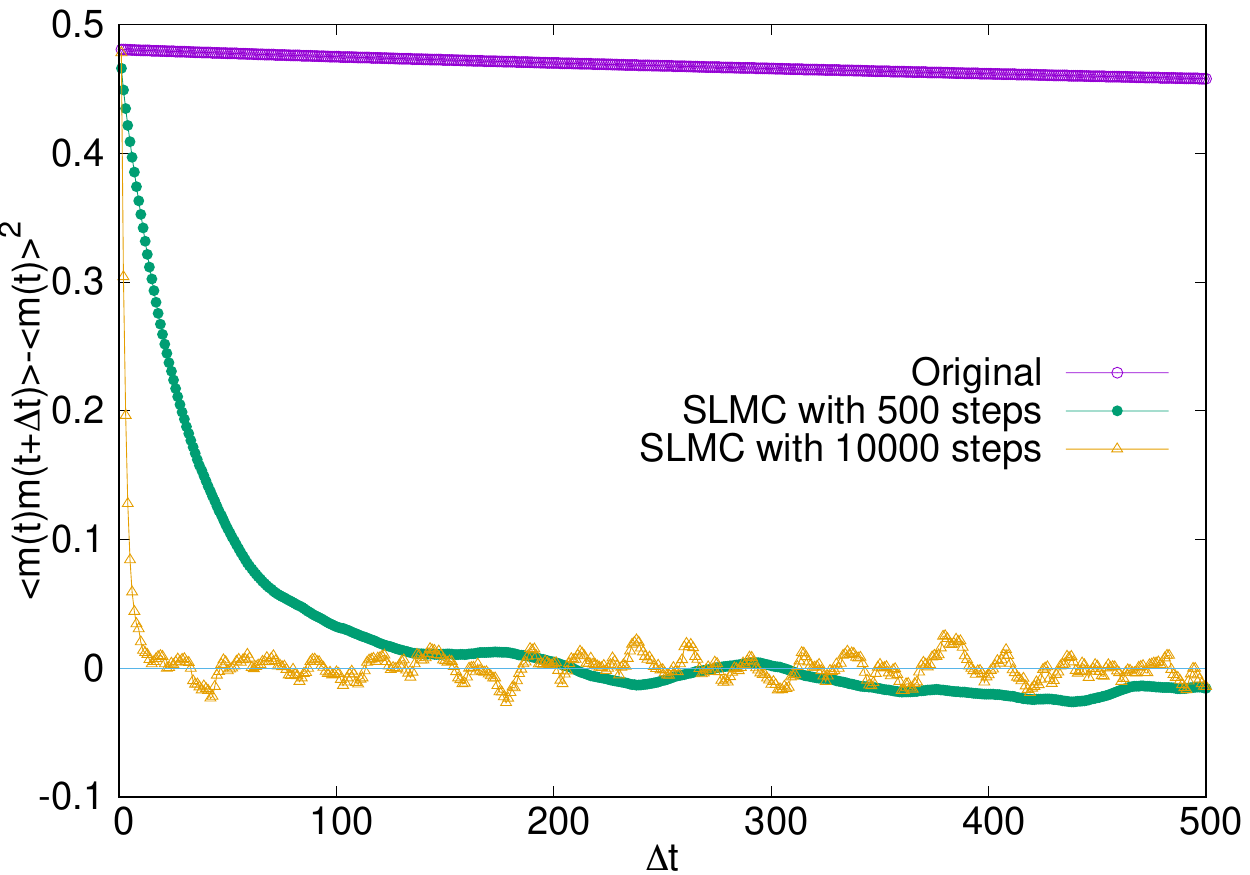}} 
    \end{tabular}
\end{center}
\caption{
\label{fig:autoc}
(Color online) 
Autocorrelation function of the auxiliary-spin magnetization for a quantum impurity model with $\beta = 40$, $U = 3D$, $\delta = 0.5$ and $V = 1D$.
Unit time is defined in the main text. 
 }
\end{figure}

To demonstrate the speedup of the SLMC, we compute the autocorrelation function of the auxiliary spin polarization defined by $m \equiv (1/N) \sum_{i=1}^N s_i$. 
Figure \ref{fig:autoc} shows the autocorrelation time of the original CT-INT method and the SLMC with the BPNN with $10$ units ($N_u = 10$), defined in terms of the number of local updates. 
We consider $\beta = 40$, $U = 3D$, $\delta = 0.5$, $V = 1D$, $m_{\rm cut} = 10$, and $n_{\rm max}=3$. 
We set the numbers for the SLMC steps $n$ are $n = 500$ and $n =10000$, whose acceptance ratios are 76\% and 63\%, respectively. 
It is clear that the fluctuation of the polarization $\langle m(t)^2 \rangle - \langle m(t) \rangle^2$ is same in all three cases. 
In this parameter region, the autocorrelation time is very long in the original CT-INT method. 
The autocorrelation function decays rapidly with the number of global moves proposed by the SLMC. 
As discussed in Sec.~\ref{sec:cost}, the autocorrelation time is expressed as $1 \sim 1/\langle A \rangle$ when the number of steps is longer than the autocorrelation time of the original simulation. 

\section{Discussion}
\subsection{Computational cost and fast updates}
We estimate the computational cost of SLMC with BPNNs.
There are two factors for evaluating this cost, the ratio of the computational cost $u_{\rm SLMC}/u_{\rm original}$ and the acceptance ratio $\langle A \rangle$, as shown in Sec.~\ref{sec:cost}. 
The computational cost to calculate the effective Hamiltonian $H_{\rm eff}({\cal C})$ is estimated as follows. 
The computational cost to calculate the element-wise effective Hamiltonian $h_{\rm eff}(\Vec{c}^j)$ with the neural networks with one hidden layer is ${\cal O}(N_u M)$ when the $M$-dimensional input vector $\Vec{c}^j$ are given. 
The computational cost to calculate the coefficients $\Vec{c}^j$ is ${\cal O}(N)$ since there are $N$ $\delta$-functions shown in Eq.~(\ref{eq:rho}). 
We can reduce this cost from ${\cal O}(N)$ to ${\cal O}(1)$ with fast local updates. 
If we consider the insertion of the vertex, the new coefficient $c_m^{j,{\rm new}}$  is calculated as $c_m^{j,{\rm new}} =c_m^{j,{\rm old}} +\phi_m(\tau_{iN+1})$, whose calculation cost is ${\cal O}(1)$.  
With use of this local update, the total computational cost to calculate the effective Hamiltonian is $u_{\rm SLMC} = {\cal O}(N_u M N)$.
Here, $N$ comes from the fact that there are $N$ input vectors $\Vec{c}^j$.
We note that the order of the computational cost of SLMC with BPNNs as a functional of $N$ is equivalent to that in SLMC in the previous paper for the CTAUX\cite{Nagai}.
Since the calculation cost of the original CTINT simulation with fast updates\cite{GullRMP,Nagai} is $u_{\rm original} = 
{\cal O}(N^2)$, the SLMC is faster than the original simulation when the acceptance ratio $\langle A \rangle$ is not too small.

\section{Conclusion}
We developed SLMC method with BPNNs, which can be considered as a general scheme to construct effective Hamiltonians with many body interactions even on continuous axis.
We demonstrated the continuous-time interaction-expansion (CTINT) SLMC with BPNNs on quantum impurity models. 
The effective Hamiltonian without any prior knowledge was obtained. 
We obtained the significant improvement of the acceptance rate with respect to the SLMC with the effective Hamiltonian using explicit expression. 
This improvement implies that obtained effective Hamiltonian of SLMC with BPNNs includes many body interaction effects, 
which is omitted in the effective Hamiltonians with the explicit forms. 
Our SLMC with BPNNs has many potential applications, since this method can accept both continuous and discrete indices of interacting particles as inputs of neural networks.

\section*{Acknowledgement}

Y. N. would like to acknowledge H. Shen, Y. Qi and L. Fu for helpful discussions and comments.
The calculations were partially performed by the supercomputing systems SGI ICE X at the Japan Atomic Energy Agency. 
This work was partially supported by
JSPS--KAKENHI Grant Numbers 18K03552 to Y.N. and 18K05208 to M.O. and the ``Topological Materials Science'' (No. 18H04228) JSPS--KAKENHI on Innovative Areas to Y.N.. 

\appendix

\section{Markov Process Monte Carlo method in Physics}\label{app:markov}
The Markov chain Monte Carlo method (MCMC) is a powerful method for an integration in high-dimensional space. 
Partition functions and physical exception values can be calculated by MCMC. 
We describe the MCMC method used in physics and point out its problems and difficulties in the following. 
\subsection{Multi-dimensional integration}
Let us consider the following multi-dimensional integration: 
\begin{align}
I = \int \cdots \int dx_1 \cdots dx_N w(x_1,\cdots,x_N) f(x_1,\cdots,x_N), 
\end{align}
where $w(x_1,\cdots,x_N)$ is a localized function in $N$-dimensional space, and $f(x_1,\cdots,x_N)$ is an arbitral function. 
In the Monte Carlo method, the integration $I$ is approximated as 
\begin{align}
I \sim \sum_{{\cal C}} f(x_1,\cdots,x_N),
\end{align}
where ${\cal C} = (x_1,\cdots,x_N)$ is randomly generated with the probability $w(x_1,\cdots,x_N)$, which is called a configuration. 

In physics, the partition function $Z$ and physical exception value $\langle A \rangle$ are expressed by the multi-dimensional integration:
\begin{align}
Z &= \int \cdots \int d\phi_1 \cdots d\phi_N e^{- S(\phi_1,\cdots,\phi_N)}, \\
\langle A \rangle &= \frac{1}{Z} \int \cdots \int d\phi_1 \cdots d\phi_N A(\phi_1,\cdots,\phi_N) e^{- S(\phi_1,\cdots,\phi_N)}.
\end{align}
For example, the physical exception value $\langle A \rangle$ in the classical Ising model on one-dimensional lattice is expressed as 
\begin{align}
\langle A \rangle &= \frac{1}{Z} \sum_{s_1,\cdots,s_N} A(s_1,\cdots,s_N) e^{- \beta E (s_1,\cdots,s_N)}, \\
&\sim \frac{1}{Z}  \sum_{{\cal C}}  A(s_1,\cdots,s_N).
\end{align}
where $Z = \sum_{s_1,\cdots,s_N} \exp(- \beta E (s_1,\cdots,s_N))$, $\beta$ is an inverse temperature, $N$ is a number of spins, $s_i = \pm 1$, $E(s_1,\cdots,s_N) = J \sum_{\langle i,j \rangle} s_i s_j$, and $\sum_{\langle i,j \rangle}$ is a summation of the nearest neighbor spins.
The configuration ${\cal C} =  (s_1,\cdots,s_N) $ is randomly generated with the Boltzmann weight $\exp \left[- \beta E (s_1,\cdots ,s_N) \right]$. 
\subsection{Design of the proposals}\label{app:proposals}
One can design various kinds of the Monte Carlo method based on Eq.~(\ref{eq:acc}). 
We need the update method from ${\cal C}$ to ${\cal C}'$ with the high acceptance ratio for good efficiency of the MCMC. 

\subsubsection{Hand-designed proposals: conventional local and global updates}
The most simple update method is so-called local update, where the configuration is updated locally. 
One randomly chooses a single site in the current configuration and proposes a new configuration by changing the variable on this site. 
For example, in the Ising spin model, the candidate of the next configuration is generated by flipping a randomly chosen spin. 
In this case, the proposal probability from $C'$ to $C$ $g({\cal C} |{\cal C}')$ equals one from $C$ to $C'$ $g({\cal C}' |{\cal C})$. 
The acceptance ratio in Eq.~(\ref{eq:acc}) becomes $A({\cal C}',{\cal C}) = {\rm min} \left(1,  \frac{w({\cal C}')}{w({\cal C})}  \right)$. 
In the local update, one can expect that $w({\cal C})$ is similar to $w({\cal C}')$ so that the acceptance rate is high, since 
the configuration ${\cal C}$ is similar to ${\cal C}'$.
Although the local update is general-purpose, model-independent method, there is an evident disadvantage. 
When  $w({\cal C}')$ is not similar to $w({\cal C})$ even if ${\cal C}'$ is similar to ${\cal C}$, it is hard to obtain uncorrelated configurations. 
Thus, the local update suffers heavily from a critical slowing down close to phase transitions. 
In such cases, the autocorrelation time within the Markov chain $\tau$ becomes very large. 

To overcome the increase of the autocorrelation time for the local update, various kinds of global update method have been developed \cite{Swendsen,Wolff,Prokofev,EvertzPRL,Evertz}. 
In all these global update methods, variables on a large number of sites are simultaneously changed in a single Monte Carlo update.  
The autocorrelation time $\tau$ can be reduced in these methods. 
The global update method is designed for specific model, since one has to use properties that a model has to obtain good efficiency. 
For example, the Wolff method can simulate the two-dimensional Ising model. 
However, by adding the interaction among the four spins in the same plaquette to this model, no simple and efficient global update method is known\cite{Liu}. 
For a given generic model, it is hard to design an efficient global update method. 

\subsubsection{Proposals by time evolution: Hybrid Monte Carlo updates}\label{app:hmc}
The hybrid Monte Carlo (HMC) method is known as the one of the good global update methods, which is widely used in the lattice quantum chromodynamics (QCD). 
In the HMC, by introducing pseudo momentum, the pseudo-time evolution from the configuration ${\cal C}$ generates the next configuration ${\cal C}'$, which is called a molecular dynamics (MD) evolution in analogy to simulations of classical particles. 
In this method, $g({\cal C} |{\cal C}')$  equals $g({\cal C}' |{\cal C})$ , since the time evolution due to the Hamiltonian dynamics
is time-reversal symmetric. 
The exact MD evolution due to the Hamilton equation conserves ``energy'' $\log w$. 
In actual simulations, the discretized time step breaks the energy conservation.
Thus, the MD evolution generates the configuration ${\cal C}'$ whose probability $w({\cal C}')$ is not equal but similar to $w({\cal C})$ so that the acceptance ratio 
$A({\cal C}',{\cal C}) = {\rm min} \left(1,  \frac{w({\cal C}')}{w({\cal C})}  \right)$ is high. 
Recently, the HMC method is revisited in the condensed matter physics to treat strongly correlated electron systems\cite{Beyl}, 
since the HMC method is general to obtain uncorrelated configurations. 
Although the HMC method might be suitable for reducing the autocorrelation time, its computational cost of the HMC is not small. 
One has to calculate the forces which is the derivative of the energy to do the MD evolution. 
For example, in the lattice QCD simulation, the inverse matrices related to the Dirac operator on a four dimensional lattice have to be calculated.

\section{Machine learning technique in Molecular Dynamics: Behler--Parrinello neural networks}\label{app:bpnn}

Machine learning techniques are widely used in the research fields of
physics and chemistry. For example, ``material informatics'' has been
developing rapidly\cite{Ramprasad2017}. 
Another remarkable development can be found in MD simulations. 
Machine learning techniques enable us to perform MD simulations in large systems
with high accuracy comparable to quantum mechanical calculation (e.g.,
density functional theory) and low computational costs close to classical MD. 
We briefly introduce the machine learning techniques in MD simulations in this section.

\subsection{Potential energy surfaces}
The basic procedure to realize large-system MD simulation is
construction of PES in a large systems by ``patching'' PESs
obtained in small systems. 
The outline is shown below. First, we
prepare a small system consisting of $N^{({\rm S})}$ atoms. Quantum
mechanical calculations are performed to obtain an accurate PES $E_{\rm
tot}^{({\rm S})} \! \left( \mathcal{R}^{({\rm S})} \right)$, where $\mathcal{R}^{({\rm S})} =
\left\{ {\bm r}_i \, | \, i = 1, \cdots, N^{({\rm s})} \right\}$ and ${\bm r}_i$ are a
set representing an atom configuration and the coordinate of the $i$-th atom, respectively.  

Next, the total energy in the small system $E_{\rm tot}^{({\rm S})} \! \left(
\mathcal{R}^{({\rm S})} \right)$ is assumed to be divided into ``partial
energies'' associated with each atoms in the system, which are
determined by ``environment'' around each atoms, i.e., 
\begin{equation}
    E_{\rm tot}^{({\rm S})} \! \left( \mathcal{R}^{({\rm S})} \right) =
    \sum_{i=1}^{N^{({\rm s})}} E_{\rm part}^{({\rm S})} \! \left( \Delta^{({\rm S})}_i
    \right) \, ,
\end{equation}
where
\begin{equation}
\Delta^{({\rm S})}_i = \left\{ {\bm r}_{ij} \, \big| \, j = 1,
\cdots, i - 1, i+1, \cdots , N^{({\rm s})} \right\} \label{delS}
\end{equation}
is a set representing an atom configuration around the $i$-th atom (${\bm r}_{ij} =
{\bm r}_j - {\bm r}_i$).

\subsection{Symmetry and descriptor}
Here, we consider energy degeneracy due to translational and rotational
invariance of the configurations, i.e., energies determined by a configuration and
translated and rotated ones have a same value. Elimination of the
degeneracy is desired for accurate evaluation of PES. Therefore, 
a set called ``descriptor'' (or ``fingerprint'')\cite{Ramprasad2017} is
introduced as 
\begin{equation}
\mathcal{F}^{({\rm S})}_i = \left\{ F_I \! \left( \Delta^{({\rm S})}_i
\right) \Big| \, I=1, \cdots, M \right\} \, , 
\end{equation}
where $F_I$ and $M$ are an function $F_I: \Delta^{({\rm S})}_i
\rightarrow \mathbb{R}$ and 
the number of them, respectively. Using the descriptor, the total energy is
given as
\begin{equation}
    E_{\rm tot}^{({\rm S})} \! \left( \mathcal{R}^{({\rm S})} \right) =
    \sum_{i=1}^{N^{({\rm s})}} E_{\rm part}^{({\rm S})} \! \left(
    \mathcal{F}_i^{({\rm S})} \right) \, ,
\end{equation}
The descriptor $\mathcal{F}_i^{({\rm S})}$ is typically a set of 
functions of relative distances between atoms and angles
between two vectors from one atom to other two atoms, which are
obviously translational and rotational invariant. For example, the
following function was proposed\cite{Behler2007,Behler2015},
\begin{align}
    F_I^{({\rm d})} \! \left( \Delta^{({\rm S})}_i \right) &= \sum_{j=1, j \ne i}^{N^{({\rm
    S})}} f^{({\rm d})}_{i,j;I} \, , \\
    f^{({\rm d})}_{i,j;I} & = \exp \left[ - \eta_I \left( r_{ij} - r_I
    \right)^2 \right] , \label{exampleFI}
\end{align}
where $r_{ij} = \left| {\bm r}_{ij} \right|$ and $\eta_I$ and $r_I$
are parameters. The function (\ref{exampleFI}) works as a detector of
bond length around $r_I$, and it is obviously translational and
rotational invariant. Here is an another example of the function\cite{Behler2007,Behler2015},
\begin{align}
    F_I^{({\rm a})} \! \left( \Delta^{({\rm S})}_i \right)
    & = \sum_{j=1, j \ne i}^{N^{({\rm S})}} \sum_{k=1 , k \ne i, j}^{N^{({\rm S})}}
    f^{({\rm a})}_{i,j,k;I} \, , \\
    f^{({\rm a})}_{i,j,k;I} &= 2^{1-\zeta_I} \left( 1 +
    \lambda_I \cos \theta_{ijk} \right)^{\zeta_I} {\rm e}^{-\eta_I
    \left( r_{ij} + r_{ik} \right)} , 
\end{align}
where $\theta_{ijk} = {\bm r}_{ij} \cdot {\bm r}_{ik} / r_{ij} r_{ik}$
and $\zeta_I$ and $\lambda_I$ are parameters. 
This function can detect the angle, and it is also translational and
rotational invariant.
These functions are
usually used as elements of a descriptor, i.e.,
\begin{equation}
    F_I = 
    \begin{cases}
        F_I^{({\rm d})} & (I = 1, \cdots, m) \\
        F_I^{({\rm a})} & (I = m+1, \cdots, M)
    \end{cases} \, .
\end{equation}

\subsection{Extension from small systems to large systems}
Toward the extension from the small systems to the large system, we introduce
a further assumption: the partial energy $E_{\rm part}^{({\rm S})}
\! \left( \mathcal{F}_i^{({\rm S})} \right)$ can
be approximately determined by the configuration of the atoms whose
distances from the $i$-th atom are less than a cutoff radius $r_{\rm c}$.
It is realized by introducing a new descriptor 
\begin{equation}
    \mathcal{G}_i^{({\rm S})} (r_{\rm c}) = \left\{ G_I \! \left(
    \Delta^{({\rm S})}_i; r_{\rm c} \right) \Big| \, I = 1, \cdots, M
    \right\} ,
\end{equation}
where the function $G_I$ satisfies the following condition:
\begin{equation}
    G_I \! \left( \tilde{\Delta}^{({\rm S})}_i; r_{\rm c} \right) = 0 \,
    , 
\end{equation}
\begin{align}
    & \tilde{\Delta}^{({\rm S})}_i (r_{\rm c}) \nonumber \\
    & = \left\{ {\bm r}_{ij} \,
    \Big| \, r_{ij} > r_{\rm c}, \, j = 1, \cdots, i-1, i+1, \cdots, N^{({\rm
    S})} \right\} \, .
\end{align}
Using the functions, we obtain the following equation.
\begin{equation}
    E_{\rm part}^{({\rm S})} \! \left( \mathcal{F}_i^{({\rm S})} \right)
    \simeq E_{\rm part}^{({\rm S})} \! \left( \mathcal{G}_i^{({\rm S})}
    \left( r_{\rm c} \right) \right) \, ,
\end{equation}
This assumption means that the partial energies are determined by not
global configurations $\Delta^{({\rm S})}_i$ but local ones
$\Delta^{({\rm S})}_i \setminus \tilde{\Delta}^{({\rm S})}_i (r_{\rm c})
= \left\{ {\bm r}_{ij} \, \big| \, r_{ij} \le r_{\rm c}, \, j =  1,
\cdots, i-1, i+1, \cdots N^{({\rm S})} \right\}$. Note that the assumption is
justified when there is no long range interaction among the atoms or
long range interactions are screened in a system.
We show an example of functions for the local descriptor below,
\begin{align}
    G_I & = 
    \begin{cases}
        G_I^{({\rm d})} & (I = 1, \cdots, m) \\
        G_I^{({\rm a})} & (I = m+1, \cdots, M)
    \end{cases} \, , \\
    G_I^{({\rm d})} \! \left( \Delta^{({\rm S})}_i; r_{\rm c} \right) &=
    \sum_{j=1, j \ne i}^{N^{({\rm S})}} f^{({\rm d})}_{i,j;I} \,
    \xi_{ij} \! \left( r_{\rm c} \right) \, , \\ 
    G_I^{({\rm a})} \! \left( \Delta^{({\rm S})}_i; r_{\rm c} \right) &=
    \sum_{j,k=1, j,k \ne i, k \ne j}^{N^{({\rm S})}} \! f^{({\rm a})}_{i,j,k;I} \, \xi_{ij} \! \left(
r_{\rm c} \right) \, \xi_{ij} \! \left( r_{\rm c} \right) ,
    \\
    \xi_{ij} \! \left( r_{\rm c} \right) & = 
    \begin{cases}
        \displaystyle
        \frac{1}{2} \left[ \cos \left(\frac{\pi r_{ij}}{r_{\rm c}}
        \right) + 1 \right] & \left( r_{ij} \le r_{\rm c} \right) \\
        0 & \left( r_{ij} > r_{\rm c} \right)
    \end{cases}
    \, .
\end{align}

Finally, we obtain the following partitioning of the total energy to the
partial energies,
\begin{equation}
    E_{\rm tot}^{({\rm S})} \! \left( \mathcal{R}^{({\rm S})} \right) =
    \sum_{i=1}^{N^{({\rm S})}} E_{\rm part}^{({\rm S})} \! \left(
    \mathcal{G}_i^{({\rm S})} \! \left( r_{\rm c} \right) \right) \, ,
\end{equation}
This partitioning enables us to extend MD simulations from the small
systems to the large system by constructing the total energy in the large system
$E_{\rm tot}^{({\rm L})}$ as
\begin{equation}
    E_{\rm tot}^{({\rm L})} \! \left( \mathcal{R}^{({\rm L})} \right) = \sum_{i =
    1}^{N^{({\rm L})}} E_{\rm part}^{({\rm S})} \! \left(
    \mathcal{G}_I^{({\rm L})} \! \left( r_{\rm c} \right) \right) \, ,
\end{equation}
where
\begin{align}
    \mathcal{R}^{({\rm L})} & = \left\{ {\bm r}_i \, \Big| \, i = 1,
    \cdots , N^{({\rm L})} \right\} \, , \\
    \mathcal{G}_i^{({\rm L})} (r_{\rm c}) & = \left\{ G_I \! \left(
    \Delta^{({\rm L})}_i; r_{\rm c} \right) \Big| \, I = 1, \cdots, M
    \right\} \, , \\
    \Delta^{({\rm L})}_i & = \left\{ {\bm r}_{ij} \, \Big| \, j = 1,
    \cdots , i-1, i+1, \cdots , N^{({\rm L})} \right\} \, ,
\end{align}
and $N^{({\rm L})}$ is the number of atoms in the large system. We
mention that the function $E_{\rm part}^{({\rm S})}$ defined in the
small system works even with the local descriptor in the large system
$\mathcal{G}_I^{({\rm L})} \! \left( r_{\rm c} \right)$. Because it
contains information of the atom configuration within the cutoff
$r_{\rm c}$, which is consistent with the local descriptor in the small
system.

\subsection{Behler--Parrinello neural networks}

Now we have a method of the extension. The remaining problem is how to
obtain the function $E_{\rm part}^{({\rm S})}$. 
Machine learning techniques give a solution of the problem.
Behler and Parrinello (BP) proposed an energy partitioning method using
neural networks \cite{Behler2007}, i.e., the
partial energies are determined by neural networks. 
Although this approach loses explicit
physical meaning of the partial energies, it is effective.

We mention that there was a serious problem about the descriptors. 
The number of the descriptors blows up as increase of the
number of atom types. 
For example, the number of descriptors is in
proportion to ${}_{N_{\rm t}}C_3$ if they contain the angles between the
centered atom and other two atoms when they have $N_{\rm t}$ types.
Artrith {\it et al.} solved the problem by introducing new descriptor
using the Chebyshev polynomials\cite{Artrith2017}. 
The new descriptor expresses information of the radial and angular distribution functions
including information of the atom types by Chebyshev coefficients. They
were succeeded in providing accurate energies of materials containing eleven atom types\cite{Artrith2017}.

%

\section{Batch-atom normalization}
We describe the detail of the batch-atom normalization in the CTINT QMC simulation. 
We consider neural networks with one hidden layer. 
The input vector is $\Vec{c}^j = (c_0^j,\cdots,c_{m_{\rm cut}-1}^j,d_0^j,\cdots,d_{m_{\rm cut}-1}^{j} )^T$.
Here, the coefficients are defined as 
\begin{align}
    c_m^j &= \sum_{i=1}^N \phi_m(\tau_{ij}), \\
    d_m^j &=  \sum_{i=1}^N s_i s_j \phi_m(\tau_{ij}),
\end{align}
with the pseudo-spin index $s_i$. 
The effective Hamiltonian is defined as 
\begin{align}
    H_{\rm eff}({\cal C}) = \frac{1}{N} \sum_{j=1} h_{\rm eff}(\Vec{c}^j) + f(N), 
\end{align}
where 
\begin{align}
    h_{\rm eff}(\Vec{c}^j) &= 
    \hat{W}_2 F(\hat{W}_1 \Vec{c}^j + \Vec{b}_1) + b_2, \\
    &= \sum_{j_1=1}^{N_u}  [\hat{W}_2 ]_{j_1} 
    F \left( [\Vec{x}]_{j_1}^j \right) + b_2,
\end{align}
where 
\begin{align}
   [\Vec{x}]_{j_1}^j \equiv \sum_{j_2=1}^M [\hat{W}_1]_{j_1 j_2}  [\Vec{c}^j]_{j_2} + [\Vec{b}_1]_{j_1}
\end{align}
Here, $[F(\Vec{x})]_{i} = F([\Vec{x}]_{i})$ denotes the activation function, $\hat{W}_{1}$ is the $N_{u} \times M$ matrix, $\hat{W}_{2}$ is $1 \times N_{u}$ matrix, 
$\Vec{b}_{1}$ is the $N_{u}$-dimensional bias vector,  $b_{2}$ is the bias, and $M$ is a number of coefficients $\Vec{c}^j$. 
We introduce the batch-atom normalization function $G$ as 
\begin{align}
    h_{\rm eff}(\Vec{c}^j) &= \sum_{j_1=1}^{N_u}  [\hat{W}_2 ]_{j_1} 
    G(F \left( [\Vec{x}]_{j_1}^j \right)) + b_2,
\end{align}
where 
\begin{align}
    G(F \left( [\Vec{x}]_{j_1}^j \right)) = \gamma_{j_1} \frac{ F \left( [\Vec{x}]_{j_1}^j \right)-\mu_{j_1}
    } {
    \sqrt{\sigma_{j_1}^{2} +  \epsilon^2}
    }+ \beta_{j_1}.
\end{align}
Here, the parameters $\gamma_{j_1}$ and $\beta_{j_1}$ are trainable parameters. $\epsilon$ is a small number. 
The batch-atom mean $\mu_{j_1}$ and variance $\sigma_{j_1}$ 
are defined as 
\begin{align}
    \mu_{j_{1}} &= \frac{1}{N N_{\rm batch}} \sum_{l=1}^{N_{\rm batch}} \sum_{j=1}^N
    F \left( [\Vec{x}^l]_{j_1}^j \right), \\
    \sigma_{j_1}^{2} &= 
    \frac{1}{N N_{\rm batch}} \sum_{l=1}^{N_{\rm batch}} \sum_{j=1}^N
    \left(F \left( [\Vec{x}^l]_{j_1}^j \right) - \mu_{j_{1}} \right)^2.
\end{align}
The index $l$ in $\Vec{x}^l$ is the index of the training data. 
Here, $N_{\rm batch}$ is the number of the batch size of the training data.


\begin{thebibliography}{99}
\bibitem{Swendsen}
R. H. Swendsen and J.-S. Wang, Phys. Rev. Lett. {\bf 58}, 86 (1987).
\bibitem{Wolff}
U. Wolff, Phys. Rev. Lett. {\bf 62}, 361 (1989).
\bibitem{Prokofev}
N. Prokof'ev, B. Svistunov, and I. Tupitsyn, Phys. Lett. A {\bf 238}, 253 (1998).
\bibitem{EvertzPRL}
H. G. Evertz, G. lana, and M. Marcu, Phys. Rev. Lett. {\bf 70}, 875 (1993)
\bibitem{Evertz}
H. G. Evertz, Adv. Phys. {\bf 52}, 1 (2003).

\bibitem{Beyl}
S. Beyl, F. Goth, and F. F. Assaad, 
Revisiting the hybrid quantum Monte Carlo method for Hubbard and electron-phonon models, 
Phys. Rev. B {\bf 97}, 085144 (2018). 


\bibitem{Blakendecler}
R. Blankenbecler, D. J. Scalapino, and R. L. Sugar,Monte Carlo
calculations of coupled boson-fermion systems. I, 
Phys. Rev. D {\bf 24}, 2278 (1981).
\bibitem{Hirsch}
 J. E. Hirsch, Two-dimensional Hubbard model: Numerical simulation study, Phys. Rev. B {\bf 31}, 4403 (1985).
\bibitem{HirschPRL}
J. E. Hirsch and R. M. Fye, Monte Carlo Method for Magnetic Impurities in Metals, Phys. Rev. Lett. {\bf 56}, 2521 (1986).
\bibitem{White}
S. R. White, D. J. Scalapino, R. L. Sugar, E. Y. Loh, J. E. Gubernatis, and R. T. Scalettar, 
Numerical study of the two dimensional Hubbard model, Phys. Rev. B {\bf 40}, 506 (1989).
\bibitem{Rubtsov}
A. N. Rubtsov, V. V. Savkin, and A. I. Lichtenstein, Continuous time
quantum Monte Carlo method for fermions, Phys. Rev. B
{\bf 72}, 035122 (2005).
\bibitem{Werner} 
P. Werner, A. Comanac, L. de Medici, M. Troyer, and A. J. Millis, Continuous-Time Solver for Quantum Impurity Models,
Phys. Rev. Lett. {\bf 97}, 076405 (2006).
\bibitem{GullRMP}
E. Gull, A. J. Millis, A. I. Lichtenstein, A. N. Rubtsov, M.
Troyer, and P. Werner, Continuous-time Monte Carlo methods
for quantum impurity models, Rev. Mod. Phys. {\bf 83}, 349 (2011).
\bibitem{GullEPL}
 E. Gull, P. Werner, O. Parcollet and M. Troyer, Continuous-time
auxiliary-field Monte Carlo for quantum impurity models,
Europhys. Lett. {\bf 82}, 57003 (2008).

\bibitem{Liu}
J. Liu, Y. Qi, Z. Y. Meng, and L. Fu, 
Self-learning Monte Carlo method, 
Phys. Rev. B {\bf 95}, 041101(R) (2017).
\bibitem{HuangWang}
L. Huang and L. Wang, 
Accelerated Monte Carlo simulations with restricted Boltzmann machines, 
Phys. Rev. B 95, 035105 (2017).
\bibitem{LiuShen}
J. Liu, H. Shen, Y. Qi, Z. Y. Meng, and L. Fu,
Self-learning Monte Carlo method and cumulative update in fermion systems, 
Phys. Rev. B {\bf 95}, 241104(R) (2017).
\bibitem{Xu}
X. Y. Xu, Y. Qi, J. Liu, L. Fu, and Z. Y. Meng, 
Self-learning quantum Monte Carlo method in interacting fermion systems, 
Phys. Rev. B {\bf 96}, 041119(R).
\bibitem{ZHLiu}
Z. H. Liu, X. Y. Xu, Y. Qi, K. Sun, Z. Y. Meng, 
Itinerant quantum critical point with frustration and non-Fermi-liquid, 
arXiv:1706.10004
\bibitem{Chen}
C. Chen, X. Y. Xu, J. Liu, G. Batrouni, R. Scalettar, and Z. Y. Meng, 
Symmetry-enforced self-learning Monte Carlo method applied to the Holstein model,
Phys. Rev. B {\bf 98}, 041102(R) (2018). 
\bibitem{HuangYang}
L. Huang, Y.-f. Yang, and L. Wang, 
Recommender engine for continuous-time quantum Monte Carlo methods, 
Phys. Rev. E {\bf 95}, 031301(R) (2017).
\bibitem{Nagai}
Y. Nagai, H. Shen, Y. Qi, J. Liu, and L. Fu, 
Self-learning Monte Carlo method: Continuous-time algorithm, 
Phys. Rev. B {\bf 96}, 161102(R) (2017). 
\bibitem{Tanaka}
A. Tanaka and A. Tomiya, 
Towards reduction of autocorrelation in HMC by machine learning, 
arXiv:1712.03893

\bibitem{Zdeborova}
L. Zdeborova, 
Machine learning: New tool in the box, 
Nat. Phys. {\bf 13}, 420 (2017).

\bibitem{Carrasquilla}
J. Carrasquilla and R. G. Melko, 
Machine learning phases of matter, 
Nat. Phys. {\bf 13}, 431 (2017).

\bibitem{Carleo}
G. Carleo and M. Troyer, 
Solving the quantum many-body problem with artificial neural networks, 
Science {\bf 355}, 602 (2017).


\bibitem{TanakaTomiya}
A. Tanaka and A. Tomiya, 
Detection of Phase Transition via Convolutional Neural Networks, 
J. Phys. Soc. Jpn. {\bf 86}, 063001 (2017).

\bibitem{ENieuwenberg}
E. P. L. van Nieuwenburg,	Y.-H. Liu, and Sebastian D. Huber, 
Learning phase transitions by confusion, 
Nat. Phys. {\bf 13}, 435 (2017).  

\bibitem{YiZhang}
Yi Zhang, and E.-A. Kim, 
Quantum Loop Topography for Machine Learning, 
Phys. Rev. Lett. {\bf 118}, 216401 (2017).

\bibitem{JChen}
J. Chen, S. Cheng, H. Xie, L. Wang, and T. Xiang, 
Equivalence of restricted Boltzmann machines and tensor network states, 
Phys. Rev. B {\bf 97}, 085104 (2018). 

\bibitem{Ramprasad2017}
R. Ramprasad, R. Batra, G. Pilania, A. Mannodi-Kanakkithodi, and C. Kim,
Machine learning in materials informatics: recent applications and
prospects, npj Comp. Mater. {\bf 3}, 54 (2017).aaa

\bibitem{Deng}
D.-L. Deng, X. Li, and S. Das Sarma, 
Quantum Entanglement in Neural Network States, 
Phys. Rev. X {\bf 7}, 021021 (2017).

\bibitem{YHuang}
Y. Huang, and J. E. Moore, 
Neural network representation of tensor network and chiral states, 
arXiv:1701.06246.

\bibitem{Hu}
W. Hu, R. R.P. Singh, R. T. Scalettar, 
Discovering Phases, Phase Transitions and Crossovers through Unsupervised Machine Learning: A critical examination, 
Phys. Rev. E 95, 062122 (2017). 


\bibitem{ZiCai}
Z. Cai, 
Approximating quantum many-body wave-functions using artificial neural networks, 
Phys. Rev. B {\bf 97}, 035116 (2018).

\bibitem{Fujita}
H. Fujita, Y. O. Nakagawa, S. Sugiura and M. Oshikawa, 
Construction of Hamiltonians by machine learning of energy and entanglement spectra, 
Phys. Rev. B {\bf 97}, 075114 (2018)

\bibitem{SJohann}
S. J. Wetzel, and M. Scherzer, 
Machine Learning of Explicit Order Parameters: From the Ising Model to SU(2) Lattice Gauge Theory, 
Phys. Rev. B {\bf 96}, 184410 (2017).

\bibitem{Shiba}
S. Iso, S. Shiba, S. Yokoo
Scale-invariant Feature Extraction of Neural Network and Renormalization Group Flow, 
Phys. Rev. E {\bf 97}, 053304 (2018). 

\bibitem{Mano}
T. Mano, and T. Ohtsuki
Phase Diagrams of Three-Dimensional Anderson and Quantum Percolation Models Using Deep Three-Dimensional Convolutional Neural Network
J. Phys. Soc. Jpn. {\bf 86}, 113704 (2017).

\bibitem{HuitaoNN}
H. Shen, J. Liu, and L. Fu, 
Self-learning Monte Carlo with deep neural networks
Phys. Rev. B 97, 205140 (2018). 

\bibitem{Bishop}
C. M. Bishop, Pattern Recognition and Machine Learning, Springer; 1st ed. 2006.


\bibitem{Blank1995}
T.B. Blank, S.D. Brown, A.W. Calhoun, and D.J. Doren, J. Chem. Phys.
{\bf 103}, 4129 (1995).

\bibitem{Brown1996}
D.F.R. Brown, 
Combining ab initio computations, neural networks, and diffusion Monte Carlo: An efficient method to treat weakly bound molecules, 
J. Chem. Phys. {\bf 105}, 7597 (1996).

\bibitem{Lorenz2004}
S. Lorenz, A. Gro\ss, and M. Scheffler, 
Representing high-dimensional potential-energy surfaces for reactions at surfaces by neural networks, Chem. Phys. Lett. {\bf 395}, 210 (2004).

\bibitem{Behler2007}
J. Behler and M. Parrinello, 
Generalized Neural-Network Representation of High-Dimensional Potential-Energy Surfaces,
Phys. Rev. Lett. {\bf 98}, 146401 (2007).

\bibitem{Behler2015}
J. Behler, 
Constructing High-Dimensional Neural Network Potentials: A Tutorial
Review,
Int. J. Quantum Chem. {\bf 115}, 1032 (2015).

\bibitem{Artrith2017}
N. Artrith, A. Urban, and G. Ceder, 
Efficient and accurate machine-learning interpolation of atomic energies in compositions with many species, 
Phys. Rev. B {\bf 96}, 014112 (2017).

\bibitem{Artrith2011}
N. Artrith, T. Morawietz, and J. Behler, High-dimensional neural-network
potentials for multicomponent systems: Applications to zinc oxide, Phys.
Rev. B {\bf 83}, 153101 (2011).

\bibitem{Artrith2014}
N. Artrith and A.M. Kolpak, Understanding the Composition and Activity
of Electrocatalytic Nanoalloys in Aqueous Solvents: A Combination of DFT
and Accurate Neural Network Potentials, Nano Lett. {\bf 14}, 670 (2014).


\bibitem{Li}
W. Li, Y. Ando, and S. Watanabe, 
Cu Diffusion in Amorphous Ta2O5 Studied with a Simplified Neural Network Potential, 
J. Phys. Soc. Jpn. {\bf 86}, 104004 (2017).

\bibitem{Liarxiv}
W. Li and Y. Ando, 
Construction of accurate machine learning force fields for copper and silicon dioxide, 
arXiv:1807.02042.

\bibitem{Assaad2007}
F. F. Assaad and T. C. Lang, Diagrammatic determinantal quantum Monte Carlo methods: Projective schemes and applications to the Hubbard-Holstein model, Phys. Rev. B {\bf 76}, 035116 (2007). 
\bibitem{Assaad}
F. F. Assaad, Continuous-time QMC Solvers for Electronic Systems in Fermionic and Bosonic Baths, in E. Pavarini, E. Koch, D. Vollhardt, and A. Lichtenstein (eds.), DMFT at 25: Infinite Dimensions: Lecture Notes of the Autumn School on Correlated Electrons 4, (Forschungszentrum Julich), ISBN 978-3-89336-953-9 (2014).
\bibitem{Luitz}
D. J. Luitz and F. F. Assaad, 
Weak-coupling continuous-time quantum Monte Carlo study of the single impurity and periodic Anderson models with 
s-wave superconducting baths, 
Phys. Rev. B {\bf 81}, 024509 (2010). 

\bibitem{ioffe2015batch}
S. Ioffe and C. Szegedy,
Batch normalization: Accelerating deep network training by reducing internal covariate shift,
arXiv:1502.03167.

\bibitem{note2}
The method how to treat many kinds of species in BP neural networks was proposed in Ref.~\cite{Artrith2017}. 
\bibitem{notedelta}
There is a relation between CTINT and CTAUX. The perturbation order becomes same when the parameter of the CTAUX has the relation: $K = U \beta (\delta^2 - 1/4 )$\cite{Assaad}. 
%

\end{thebibliography}
\end{document}